\documentclass[nofootinbib]{revtex4-2}
\usepackage[margin=1in]{geometry}
\usepackage[utf8]{inputenc}
\usepackage{amsmath}
\usepackage{amsfonts}
\usepackage{amssymb}
\usepackage{amsthm}
\usepackage{mathtools}
\usepackage{graphicx}
\usepackage[braket,qm]{qcircuit}

\newtheorem{theorem}{Theorem}
\newtheorem{lemma}{Lemma}

\newcommand{\br}{\vspace{3mm}}

\newcommand{\vd}{\raisebox{2.5mm}{\vdots}}

\begin{document}

\title{On the Exponential Sample Complexity of the\\ Quantum State Sign Estimation Problem}

\author{Arthur G. Rattew$^{1, 2}$ and Marco Pistoia$^1$}
\affiliation{$^1$JPMorgan Chase Bank, N.A., Future Lab for Applied Research and Engineering\\ $^2$University of Oxford, Department of Materials, Oxford, United Kingdom}

\begin{abstract}
We demonstrate that the ability to estimate the relative sign of an arbitrary $n$-qubit quantum state (with real amplitudes), given only $k$ copies of that state, would yield a $kn$-query algorithm for unstructured search. Thus the quantum sample complexity of sign estimation must be exponential: $\Omega(2^{n/2}/n)$. In particular, we show that an efficient procedure for solving the sign estimation problem would allow for a polynomial time solution to the NP-complete problem 3-SAT.
\end{abstract}

\maketitle

\section{Introduction}

Reading out the relative sign of an arbitrary quantum state in a superposition is an important subroutine in a number of problems, such as tax-aware portfolio construction in finance, and in a quantum implementation of the simplex method~\cite{moehle2021tax, kerenidis2020quantum, nannicini2021fast}. For example, in tax-aware portfolio construction, one aims to solve a minimization problem where the tax-liability function is non-convex for certain inputs. A standard approach to solve this is to approximate the non-convex component with a convex function, and then to solve the resulting semi-definite program. The signs of the output of this approximate solution are then used to constrain the input to the original problem, guaranteeing its convexity and thereby allowing for an efficient solution. Thus, proving a lower bound on the efficiency of a sign estimation procedure also places a lower bound on the performance of any algorithms utilizing it as a subroutine.  

\br 

In 1996, Grover published a seminal paper presenting a quantum algorithm for unstructured search with complexity $O(\sqrt{N})$, where $N$ is the length of the list~\cite{grover1996fast}. It was then proven that in the query model, Grover search is asymptotically optimal for unstructured search~\cite{zalka1999grover}. Unstructured search, subsequently generalized and termed ``amplitude amplification'', is fundamental to many quantum algorithms that exhibit a theoretical polynomial quantum speedup~\cite{brassard2002quantum}. Of relevance, through the creation of an oracle marking the states corresponding to the satisfying solutions of an $n$-variable instance of 3-SAT, unstructured search may be directly applied to solve 3-SAT with complexity scaling as $O(\sqrt{2^n})$ (representing a square-root speedup over brute-force search). 

\br 

In this document, we prove that an efficient\footnote{As a note of terminology, when we say ``efficient'', we mean ``with polynomial complexity''.} sign estimation procedure receiving $kn$-copies of an input state would allow for a polynomial time solution to 3-SAT, the Boolean satisfiability problem where where each clause is limited to at most three literals. Moreover, we also show a lower bound on the query complexity of quantum state sign estimation. This result is particularly useful in guiding research efforts for designing new quantum algorithms that utilize a sign estimation subroutine.

\section{Main Result}
We are given an oracle $O_f$ marking the desired state $\ket{m}$ as
    \begin{align}
        O_f\ket{x} = 
        \begin{cases}
            \ket{x} & \text{if } x \neq m\\
            -\ket{x} & \text{if } x = m
        \end{cases}.
    \end{align}
We use the convention that $f(x\neq m)=1$ and $f(x=m)=-1$.
We also define the general quantum state as follows: $\ket{\psi}$,
\begin{align}
    \ket{\psi} 
    = \sum_{x = 0}^{N - 1} \alpha_x\ket{x},
\end{align}
such that $\sum_{x}|\alpha_x^2| = 1$, and $\alpha_x \in \mathbb{R}$ for all $x$.
Consider some operator $S_x$, where the index is the standard basis label of the state whose amplitude is being queried. Without loss of generality, we assume that $S_x$ is a unitary transformation which outputs $\ket{0}$ on the first qubit if $\alpha_x$ is positive, and $\ket{1}$ on the first qubit if $\alpha_x$ is negative (we can assume that $\alpha_x\neq 0$). Additionally, we suppose that $S_x$ acts on $k$-copies of the input state being queried.
The value of the other qubits are irrelevant, and they need not even be measured.
So long as $S_x$ is implemented efficiently, it does not matter if it is unitary, non-unitary, or a combination of quantum and classical, requires some polynomial number of additional ancillary qubits, or destroys the quantum state it acts upon---all that matters is that given $k$-copies of a quantum state in superposition, $S_x$ somehow returns the sign of the specified basis vector. It is worth noting that our proof regarding the lower-bound query complexity of the sign estimation procedure does not actually require $S_x$ to be efficiently implementable.
Of course, an overall global phase on a quantum state is not experimentally determinable, so we are clearly referring to the relative phase of a state.

\begin{figure}[t]
    \[
        \hspace{6em}\Qcircuit @C=1em @R=0.7em {
        &&\lstick{\ket{+}}   & \qw      & \qw & \multigate{4}{O_f} & \qw & \qw      & \qw  & \qw      & \qw & \multigate{4}{H^{\otimes n}} & \qw & \qw  \\
        &&                   &          &         &     &     &          &      &          &     &         &  &      &     &  &         &  &   & \\
        &&                   & \vd      & & \ghost{O_f}            & \qw & \vd      &      & \ctrl{3} & \qw & \ghost{H^{\otimes n}}        & \qw & \qw  \\
        &&                   &          &     &         &     &          &      &          &     &         &  &      &     &  &         &  &   & \\
        &&\lstick{\ket{+}}   & \qw      & \qw & \ghost{O_f}        & \qw & \qw      & \qw  & \qw      & \qw & \ghost{H^{\otimes n}}        & \qw & \qw  \\
        &&\lstick{\ket{0}}   & \gate{H} & \qw & \ctrl{-1}          & \qw & \gate{H} & \qw  & \gate{Z} & \qw & \ctrl{-1}                    & \qw & \qw  \inputgroupv{1}{5}{1em}{2.4em}{n}
        }
    \]
    \caption{Quantum circuit producing the state used as input to the sign estimation procedure to determine the $j^{th}$ bit, $m_j$, of the secret marked state $m$. Note that the controlled-$Z$ gate is controlled on the $j^{th}$ qubit in the main register.}
    \label{fig:mcmr_based_circuit}
\end{figure}
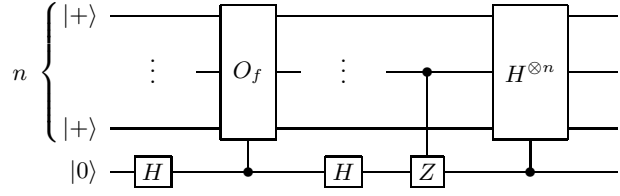

\begin{lemma}
Given an oracle $O_f$ marking the state $\ket{m}$, the quantum state,
\begin{align}
    \ket{\phi_0} = \sqrt{\frac{1}{N}}\sum_{x\neq m}\ket{x}\ket{0} + \sqrt{\frac{1}{N}}\ket{m}\ket{1}
\end{align}
may be produced with a polynomial depth circuit.
\end{lemma}
\noindent \textbf{Proof.} Let $[A]_b^a$ represent the application of the unitary $A$ on register (or qubit) $b$ conditioned on the value of register $a$. 
In our procedure, we use an $n$-qubit main register, and an ancillary register with only a single qubit. By convention, we write our states as the tensor product of two states, $\ket{a}_n\ket{b}$, where the first ket denotes the state of the $n$-qubit register, while the second ket denotes the state of the ancillary qubit.
Consider the following sequence of transformations on the starting state $\ket{+}_n\ket{0}$,
\begin{equation}
    I^{\otimes n}\otimes H\ket{+}_n\ket{0} = \ket{+}_n\ket{+}
\end{equation}
\begin{equation}
    \xrightarrow{[O_f]_{1:n}^{n+1}} \frac{1}{\sqrt{2}}\ket{+}_n\ket{0} + \left(\frac{1}{\sqrt{2N}}\sum_{x}(-1)^{f(x)}\ket{x}_n \right)\ket{1}
\end{equation}
\begin{equation}
    \xrightarrow{I^{\otimes n}\otimes H} 
    \frac{1}{2}\ket{+}_n\ket{0} + \frac{1}{2}\ket{+}_n\ket{1} + \left(\frac{1}{2\sqrt{N}}\sum_{x}(-1)^{f(x)}\ket{x}_n \right)\ket{0} - \left(\frac{1}{2\sqrt{N}}\sum_{x}(-1)^{f(x)}\ket{x}_n \right)\ket{1}
\end{equation}
\begin{equation}\label{eqn:hopeful_equation}
    = \frac{1}{\sqrt{N}}\sum_{x \neq m}\ket{x}_n \ket{0} + \frac{1}{\sqrt{N}}\ket{m}_n\ket{1}.
\end{equation}
Thus, the complexity of this procedure is simply the complexity of implementing the oracle, and is thus polynomial.\qed

\begin{lemma}\label{lemma:bit_readout}
Let $\ket{m}=\ket{m_1, ..., m_n}$.
Given, the state
\begin{align}
    \ket{\phi_0} = \sqrt{\frac{1}{N}}\sum_{x\neq m}\ket{x}_n\ket{0} + \sqrt{\frac{1}{N}}\ket{m}_n\ket{1},
\end{align}
and the operator $S_x$, we can read out the value of bit $m_j$ with polynomial complexity.
\end{lemma}
\noindent \textbf{Proof.} We can determine the value of bit $m_j$ with the following procedure. 
First, obtain
\begin{align}
    [Z]_{n+1}^{j}\ket{\phi_0}
    =
    \sqrt{\frac{1}{N}}\sum_{x\neq m}\ket{x}_n\ket{0} + (-1)^{m_j}\sqrt{\frac{1}{N}}\ket{m}_n\ket{1}.
\end{align}
Next, apply $[H^{\otimes n}]_{1:n}^{n+1}$ (i.e., apply a Hadmard gate on each qubit conditioned on the state of the ancilla qubit), as follows:
\begin{align}
    [H^{\otimes n}]_{1:n}^{n+1}[Z]_{n+1}^{j}\ket{\phi_0}
    =
    \sqrt{\frac{1}{N}}\sum_{x\neq m}\ket{x}_n\ket{0} + \sqrt{\frac{1}{N}}\sum_{x}(-1)^{m_j}(-1)^{x\cdot m}\ket{x}_n\ket{1},
\end{align}
where $x\cdot m \equiv \sum_{i=1}^n x_im_i$. 
Observing that for $x=0$, $x\cdot m = 0$ for any value of $m$, the amplitude on the state $\ket{0}_n\ket{1}$ is thus given by $\frac{(-1)^{m_j}}{\sqrt{N}}$. As a result, we wish to apply our sign-reading operator to the basis state with label $z=00\cdots 01$, obtaining the $k(n+1)$-qubit state,
\begin{align}
    \ket{\phi_1} = S_{z} \left( [H^{\otimes n}]_{1:n}^{n+1}[Z]_{n+1}^{j}\ket{\phi_0} \right)^{\otimes k}.
\end{align}
By formulation, $S_z$ will place the first qubit in the state $\ket{0}$ if the sign on state with label $z$ is $+1$, and in the state $\ket{1}$ if the sign on the state is $-1$. As a result, we can directly deduce that 
\begin{align}
    m_j
    =
    \begin{cases}
        0 & \text{if qubit 1 of } \ket{\phi_1} \text{ is in state } \ket{0}\\
        1 & \text{if qubit 1 of } \ket{\phi_1} \text{ is in state } \ket{1}
    \end{cases}.
\end{align}
Up to the application of $S_z$, given the input state $\ket{\phi_0}$ all the steps in this procedure can clearly be implemented with complexity bounded by $O(n)$. By assumption, if $S_x$ is implemented efficiently, then this procedure allows for the deterministic determination of the value of bit $m_j$ with polynomial complexity. 
The quantum circuit producing the input to the sign estimation procedure used to determine the value of bit $m_j$ is shown in Figure~\ref{fig:mcmr_based_circuit}. \qed

\begin{theorem}
For an $n$-qubit state, a quantum state sign estimation procedure requiring only $k$-copies of the input state would yield a $kn$-query algorithm for unstructured search. 
\end{theorem}

\noindent \textbf{Proof.} In Lemma~\ref{lemma:bit_readout}, we show that the ability to efficiently perform quantum state sign estimation allows the determination of the value of bit $m_j$ of the state marked by oracle $O_f$ with only $k$ queries to the oracle. This procedure may then be repeated $n$ times, once per bit in $m$, to read out the entire marked state, resulting in a total query complexity of $kn$. Moreover, if the sign estimation procedure has polynomial complexity in its implementation, then it immediately follows that Lemma~\ref{lemma:bit_readout} may be used to perform quantum unstructured search with polynomial complexity.
As a result, this polynomial time unstructured search procedure could be immediately used to efficiently find satisfying solutions to 3-SAT, thus proving that quantum state sign estimation must be at least as hard as the NP-complete problem 3-SAT. \qed

\br

As a corollary, the quantum sample complexity of sign estimation must be exponential: $\Omega(2^{n/2}/n)$, following immediately from the optimality of Grover search~\cite{zalka1999grover}.

\section{Conclusion}
For an $n$-qubit system and a sign estimation procedure requiring $k$-copies of the input state, we have presented a $kn$-query algorithm for unstructured search. This proves the lower bound $\Omega(2^{n/2}/n)$ for the sample complexity of sign estimation.
Finally, we have shown that an efficient procedure for solving the sign estimation problem would immediately allow for an efficient solution to the NP-complete problem 3-SAT.

\section*{Acknowledgements}
Thanks to Shouvanik Chakrabarti, Pierre Minssen, and Dylan Herman for insightful conversations. 

\section*{Disclaimer}
This paper was prepared for information purposes by the Future Lab for Applied Research and Engineering (FLARE) group of JPMorgan Chase Bank, N.A..  This paper is not a product of the Research Department of JPMorgan Chase \& Co. or its affiliates.  Neither JPMorgan Chase \& Co. nor any of its affiliates make any explicit or implied representation or warranty and none of them accept any liability in connection with this paper, including, but limited to, the completeness, accuracy, reliability of information contained herein and the potential legal, compliance, tax or accounting effects thereof.  This document is not intended as investment research or investment advice, or a recommendation, offer or solicitation for the purchase or sale of any security, financial instrument, financial product or service, or to be used in any way for evaluating the merits of participating in any transaction.

\bibliographystyle{unsrt}
\bibliography{bibliography}

\end{document}